\documentclass[10pt,twocolumn,letterpaper]{article}

\usepackage[pagenumbers]{cvpr} 

\usepackage{algorithm}
\usepackage{algorithmic}
\usepackage{marvosym}
\usepackage{wasysym}
\usepackage{multirow}
\usepackage{multicol}
\definecolor{cvprblue}{rgb}{0.21,0.49,0.74}
\usepackage[pagebackref,breaklinks,colorlinks,allcolors=cvprblue]{hyperref}

\def\confName{CVPR}
\def\confYear{2026}

\title{FBA$^2$D: Frequency-based Black-box Attack for AI-generated Image Detection}

\author{
Xiaojing Chen, 
Dan Li, 
Lijun Peng, 
Jun Yan\Letter, 
Zhiqing Guo, \\
Junyang Chen, 
Xiao Lan, 
Zhongjie Ba, 
Yunfeng Diao\Letter
}

\begin{document}
\maketitle
\begin{abstract}
The prosperous development of Artificial Intelligence-Generated Content (AIGC) has brought people's anxiety about the spread of false information on social media. Designing detectors for filtering is an effective defense method, but most detectors will be compromised by adversarial samples. Currently, most studies exposing AIGC security issues assume information on model structure and data distribution. In real applications, attackers query and interfere with models that provide services in the form of application programming interfaces (APIs), which constitutes the black-box decision-based attack paradigm. However, to the best of our knowledge, decision-based attacks on AIGC detectors remain unexplored. In this study, we propose \textbf{FBA$^2$D}: a frequency-based black-box attack method for AIGC detection to fill the research gap. Motivated by frequency-domain discrepancies between generated and real images, we develop a decision-based attack that leverages the Discrete Cosine Transform (DCT) for fine-grained spectral partitioning and selects frequency bands as query subspaces, improving both query efficiency and image quality. Moreover, attacks on AIGC detectors should mitigate initialization failures, preserve image quality, and operate under strict query budgets. To address these issues, we adopt an ``adversarial example soup'' method,  averaging candidates from successive surrogate iterations and using the result as the initialization to accelerate the query-based attack. The empirical study  on the Synthetic LSUN dataset and GenImage dataset demonstrate the effectiveness of our prosed method. This study shows the urgency of addressing practical AIGC security problems.
\end{abstract} 
\section{Introduction}

The explosion of AIGC technology has greatly enriched people's entertainment. This technology relies on generative models, such as Generative Adversarial Networks (GANs)~\cite{NIPS2014_f033ed80} and diffusion models~\cite{NEURIPS2020_4c5bcfec}, which can generate realistic content. However, synthetic content can be exploited for illegal purposes, such as the dissemination of false information, raising public concern over the potential misuse of these technologies.

Ensuring the integrity of digital media requires reliable detection of AIGC, a challenge that has attracted significant interest in both academia and industry. State-of-the-art detectors ~\cite{Wang_2020_CVPR,zhong2024patchcraftexploringtexturepatch} predominantly leverage deep neural networks to discriminate between authentic and generated examples, demonstrating strong performance across diverse datasets and generative models. Compared with deepfake detection~\cite{wang2024deepfake}, which is a vertical, person-centric forensics task emphasizing temporal cues and physiological consistency, AIGC detection focuses more on cross-modal forensics, frequency-domain evidence, and semantic-consistency checks.

These DNN-based detectors remain vulnerable to adversarial perturbations—imperceptible, carefully crafted modifications that can mislead the classifier ~\cite{goodfellow2015explainingharnessingadversarialexamples}. Existing methods have presented the adversarial vulnerability of AIGC detectors~\cite{saberi2024robustnessaiimagedetectorsfundamental,diao2025vulnerabilitiesaigeneratedimagedetection,mavali2025adversarialrobustnessaigeneratedimage}, but these methods require access to the detector's network architecture or dataset. However, in practical applications, attackers cannot obtain this prior knowledge and will only receive a detection output. In this black-box threat model, the adversary can only query the hard label output of the classifier, having no access to its architecture, weights, confidence scores, or gradients. This attack is known as a decision-based attack. To the best of our knowledge, the research on decision-based attacks against AIGC detectors is insufficient. Unlike the multi-classification task of image classification, AIGC detection is usually defined as a binary classification problem to distinguish generated content from real content. Existing decision-based attacks usually use random noise as attack initialization, which is easy to succeed in image classification tasks. Nevertheless, the real-fake binary classification in AIGC detection is different from the common and standard cat and dog binary classification in that the binary classification of AIGC detection is asymmetric~\cite{2024arXiv241115633Y}, that is, the characteristics of the generated content only exist in a small decision area. It is difficult to initialize in this small decision area using random noise as attack initialization. The adversary can overcome this difficulty by launching a direct targeted attack. But it suffers from degraded image quality and an excessive number of queries. In practical application scenarios, the number of queries is usually limited. Therefore, it is significant to explore the decision-based black-box attack method against AIGC detectors. Furthermore, previous studies have shown that the distribution of generated images and real images in the frequency domain is different~\cite{pmlr-v119-frank20a,Chandrasegaran_2021_CVPR}. Compared with real images, the information of generated images is more concentrated in the low-frequency area, and the information in the high-frequency area is sparse.

\begin{figure}[!t]    
    \centering    
    \includegraphics[width=1\linewidth]{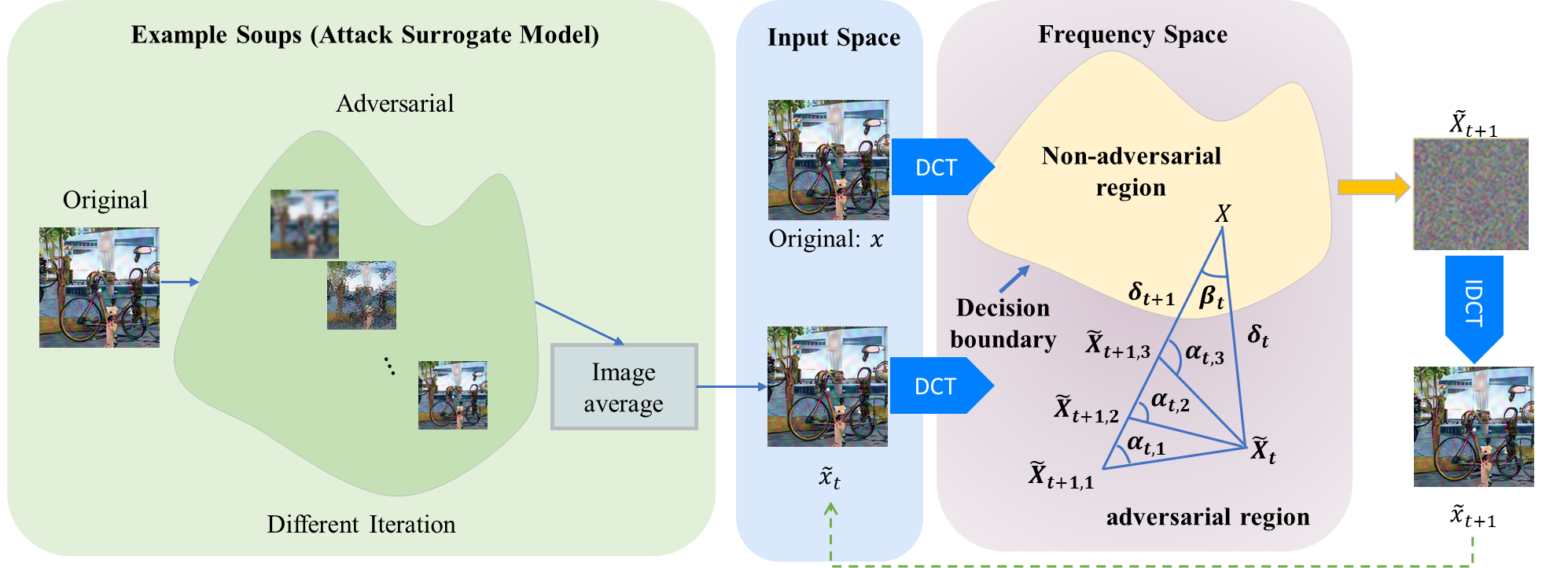}    
    \caption{Our proposed FBA$^2$D framework.}    
    \label{fig:da2d}
\end{figure}

To fill such a research gap, we propose an efficient frequency-based black-box adversarial attack method to expose the vulnerability of the AIGC detectors. Figure \ref{fig:da2d} illustrates our proposed framework. Specifically, the AI-generated images usually exhibit unnatural energy distribution patterns in high frequency components~\cite{dong2022think} with the specific fingerprints~\cite{marra2019gans}. Thus, we select different strategies for real and generated images due to their frequency domain characteristics. For real images, we employ a mixed high- and low-frequency band as the query subspace. For generated images, we restrict the query subspace to the low-frequency band, which can improve query efficiency and image quality. To improve the efficiency of the query-based attack, we use the ``example soup'' method~\cite{10858076} to obtain the average adversarial examples of the surrogate model at different iterations and query whether this average adversarial example is adversarial when attacking the target model. If the sample is adversarial, we use the averaged adversarial example as the initial perturbation of the decision-based attack. If the average example is not adversarial, we instead use the targeted attack method and employ data from another category as the initial perturbation to ensure the success of initialization. In the empirical study, our method achieves state-of-the-art attack performance and query efficiency on the Synthetic LSUN dataset~\cite{Wang_2020_CVPR} and the GenImage benchmark~\cite{zhu2023genimage}, further providing some inspiration for the security performance of AIGC detectors in real scenarios. Unlike DeepFake detectors, which primarily rely on facial symmetry, temporal coherence, or physiological consistency~\cite{wang2024deepfake}, AIGC detectors are often sensitive to distinct high-frequency noise patterns, color statistics, and texture cues arising from generative models. Our proposed method is the first decision-based black-box attack method specifically tailored to this new AIGC detection task, featuring a novel frequency-aware perturbation scheme and optimization strategy. Our main contributions are summarized as follows.
\begin{itemize}
\item To the best of our knowledge, this study proposes the first decision-based attack method for AIGC models, which adapts to the security assessment requirements of real detection scenarios.
\item Our experimental results show that real images contain low-frequency and high-frequency components, and generated images mainly contain low-frequency components.
\item Building on our empirical insights in the frequency domain, we further refine our decision-boundary attacks. For real images, we define the query subspace as the combined low- and high-frequency components. In contrast, for generated images, our black-box attack restricts queries to the low-frequency subspace, improving both query efficiency and the quality of the crafted adversarial examples.   
\end{itemize}


\section{Related Work}
\subsection{Decision-based Attacks}
In the scenarios of black-box adversarial attacks, the assaulter would not require access to the internal structure of the target model to carry out attacks, including transfer-based~\cite{Dong_2018_CVPR}, score-based~\cite{pmlr-v97-guo19a,liu2018signsgd}, and decision-based~\cite{Brendel2018DecisionBasedAA,cheng2018query,Rahmati_2020_CVPR,9152788} attacks. These studies are closely related to the real-world security issues.

Brendel et al.~\cite{Brendel2018DecisionBasedAA} proposed the first decision-based attack method, known as Boundary Attack, in which the attacker iteratively adjusts examples to maintain adversarial properties while moving toward the decision boundary. Subsequent works such as GeoDA~\cite{Rahmati_2020_CVPR} and Triangle Attack~\cite{10.1007/978-3-031-20065-6_10} provide the geometric information to explore the decision boundary, thus improving the efficiency of the attack. The OPT method~\cite{cheng2018query} leverages the zeroth-order optimization method and demonstrates strong performance in reducing query complexity. The Sign-OPT method efficiently estimates the gradient direction, accelerating the process of black-box attacks~\cite{cheng2020sign}. The SignFlip method~\cite{chen2020boosting} introduces noise projection and random sign flipping, significantly improving decision-based attacks. The HSJA method~\cite{9152788} uses the geometric properties of high-dimensional spaces and performs well in complex models. The RayS method~\cite{10.1145/3394486.3403225} reformulates the continuous problem of finding the nearest decision boundary into a discrete one, eliminating the need for zeroth-order gradient estimation. While these methods are primarily used for image classification tasks, their effectiveness in AIGC security has not been fully explored.

\subsection{Adversarial Robustness of AIGC Detectors}
AIGC detection is a security-sensitive task, which is vulnerable to adversarial examples~\cite{diao2025vulnerabilitiesaigeneratedimagedetection,saberi2024robustnessaiimagedetectorsfundamental,mavali2025adversarialrobustnessaigeneratedimage}. Early studies have revealed that GAN-generated images have artifact features in frequency domains~\cite{NEURIPS2020_1f8d87e1, pmlr-v119-frank20a}. To this end, a variety of high-accuracy frequency domain detection methods~\cite{pmlr-v119-frank20a,Chandrasegaran_2021_CVPR} have been proposed to resolve this problem. Dong et al.~\cite{Dong_2022_CVPR} demonstrated the reliability limitations of frequency artifact-based detection methods. Research on adversarial attacks against detectors has evolved in multiple dimensions. Diao et al.~\cite{diao2025vulnerabilitiesaigeneratedimagedetection} conducted the first systematic evaluation of AIGI detector vulnerabilities in both white-box and transfer-based black-box scenarios. Saberi et al.~\cite{saberi2024robustnessaiimagedetectorsfundamental} established robustness benchmarks for digital watermarking and classifier-based detection methods. In terms of attack techniques, Lee et al.~\cite{Lee_Jung_Seo_2024} proposed the Spectrum Translation for Refinement of Image Generation (STIG) framework, which enhances the stealthiness of adversarial examples through spectral transformation optimization of generated images. Zhou et al.~\cite{zhou2024stealthdiffusionevadingdiffusionforensic} proposed a StealthDiffusion method, which effectively evades state-of-the-art forensic detectors. 
Our study launches a decision-based attack against the AIGC detector to provide insight into the security of AI content in real-world applications.
\section{Method}
\label{sec:method}
In this study, we propose \textbf{FBA$^2$D}: a decision-based attack method for AI-generated image detection. Fig.\ref{fig:da2d} illustrates our proposed method. This method is inspired by the milestone work~\cite{10.1007/978-3-031-20065-6_10}. We scale the Triangle Attack paradigm to the AIGC detection scenario with a new frequency-domain signal adjustment mechanism. The clean data $x$ would be attacked as adversarial examples $\tilde{x}_{t},\ \tilde{x}_{t+1}$ in the current and next queries. In the frequency space, we apply the DCT to adjust the signals, where the variables $\widetilde{X}_{t},\ \widetilde{X}_{t+1,1},\ \widetilde{X}_{t+1,2},\ \widetilde{X}_{t+1,3}$ denote the adversarial data of different queries $t$ in the frequency domain,  and the variables $\delta_{t},\ \delta_{t+1}$ represent the perturbations. The variables $\alpha_{t, 1},\ \alpha_{t, 2},\ \alpha_{t, 3},\ \beta_{t}$ represent the search angles in the decision boundary. To improve the efficiency of the attack, we introduce the concept of ``Adversarial Example Soups'' (AES). The Projected Gradient Descent (PGD) method is applied to configure a successful initialization with iterations $I_{1},\ ...,\ I_{n}$ to transform $n$ original natural samples into adversarial examples for the average operation. The attack on the surrogate model further boosts the attack effectiveness and efficiency of the decision-based attack on the AIGC detector.

\subsection{Preliminaries}
We define an AI-generated content detector $f:\mathbb{R}^d\to\mathbb{R}^k$ as the detection model. Because AIGC detection is a binary classification task, its label $y$ is real or fake. The formulation is $y\in\mathbb{R}^{2}=\{0,1\}$. Given an original image $x$, we define $f(x)$ as the detector function and $c(x)=\mathrm{arg~max}\ f(x)$ as the predicted label.
In the decision-based attack, the adversary has no access to the network architecture, weights, gradients, or predicted probabilities of the model $f(x)$, only the predicted labels of the model $c(x)$. Therefore, the optimization objective of the decision-based attack is as follows:
\begin{equation}\min_\delta||\delta||_{p}\quad\mathrm{s.t.}\quad c(x+\delta)\neq c(x),\end{equation}
where the adversary keeps querying the model and optimizing adversarial perturbations $\delta$ until $||\delta||_{p}\leq\epsilon$, and $\epsilon$ is the allowed maximum perturbation.

\subsection{Frequency Perspective on Adversarial Robustness in AIGC Detection}
Recent advances in AIGC detection have systematically characterized the pronounced frequency domain disparities between real and AI-generated~\cite{pmlr-v119-frank20a,Chandrasegaran_2021_CVPR}, with state-of-the-art detectors achieving discrimination by capitalizing on these spectral signatures~\cite{Wang_2020_CVPR}. Most of these studies find that, compared to generated images, real images contain richer high-frequency components. Motivated by these findings, we propose a decision-based attack method to fully exploit the frequency-dependent characteristics of generated versus real images across different model architectures.

The baseline method we employ unifies geometric optimization paradigms with signal frequency-domain analysis~\cite{10.1007/978-3-031-20065-6_10} to achieve multi-band perturbation synthesis through spatial iterations and frequency-domain transformations. This method treats the input space as a signal field, deconstructing adversarial perturbations using spectral tools. The baseline is a query-efficient decision-based adversarial attack method that has good results on classification tasks. However, previous studies have shown that the binary classification problem of AIGC detection is not the same as the traditional cat-dog binary classification problem. The binary classification of AIGC detection is asymmetric~\cite{2024arXiv241115633Y}, that is, the probability of being classified into one category may be higher than the probability of being classified into the other category. The decision-based attack mechanism~\cite{10.1007/978-3-031-20065-6_10} that works in image classification cannot be directly applied to AIGC detection tasks. Specifically, the initialization of the decision-based attack fails, resulting in an attack success rate close to 0\%. It shows that random initialization is unsuccessful and cannot guarantee adversarial behavior. Therefore, we use a targeted attack to ensure a successful initialization.

In the previous classification task, the attack is built on the postulation that the image gradient estimation was concentrated in the low-frequency domain. Thus, the adversary would use the low-frequency frequency domain (the upper 10\% frequency components) as the query subspace to improve the query efficiency. When facing the AIGC detection task, the frequency domain difference between the generated image and the real image will lead to a difference in the attack paradigm. Consequently, we partition the frequency spectrum into distinct bands and evaluate, within each subspace, the quality and attack success rate of adversarial examples crafted from real versus generated images.

\begin{algorithm}[!t]
\caption{Decision-based Attack for AI-generated Image Detection}
\label{alg:algorithm}
\textbf{Input}: Detection model $f$; Original data $x$ with ground-truth label $y$; target adversarial example $x_t$ from a category different from that of the original data $x$; Maximum number of queries $Q$; Maximum number of iteration $N$ for each sampled subspace; Dimension of the directional line $d$; Lower bound $\underline{\beta}$ for angle $\beta$.\\
\textbf{Parameter}: The learned angle $\alpha$, and the search angle $\beta$.\\
\textbf{Output}: An adversarial example $\tilde{x}$. \\
\begin{algorithmic}[1] 
\STATE Initialize an adversarial example $\tilde{x}_0=x_t$;
\STATE$~\tilde{X}_0=\mathrm{DCT}(\tilde{x}_0)$, $X=\mathrm{DCT}(x)$, $q=0$, $t=0$, $\alpha_0=\pi / 2$;
\WHILE {$q<Q$}
\STATE Sample 2-D subspace $\mathcal{S}_t$ in the frequency-domain;
\STATE $\beta_{t, 0}=\max (\pi-2 \alpha_t, \underline{\beta}) $;
\IF {$c(\mathcal{T}(X, \tilde{X}_t, \alpha_{t, 0}, \beta_{t, 0}, \mathcal{S}_t))=c(X)$}
\STATE $q=q+1$;
\STATE Update $\alpha_{t, 0}$ based on Eq. (\ref{con:eq5});
\IF {$c(\mathcal{T}(X, \tilde{X}_t, \alpha_{t, 0}, -\beta_{t, 0}, \mathcal{S}_t))=c(X)$}
\STATE $q=q+1$;
\STATE Update $\alpha_{t, 0}$ based on Eq. (\ref{con:eq5});
\STATE Go to the next loop iteration;
\ENDIF
\ENDIF
\STATE $\bar{\beta}_{t, 0}=\min (\pi / 2, \pi-\alpha_t)$;
\FOR{$i=0 \rightarrow N$}
\STATE $\beta_{t, i+1}=(\bar{\beta}_{t, i}+\beta_{t, i}) / 2$;
\IF {$c(\mathcal{T}(X, \tilde{X}_t, \alpha_{t, i}, \beta_{t+1, i}, \mathcal{S}_t))=c(X)$}
\STATE $q=q+1$;
\STATE Update $\alpha_{t, 0}$ based on Eq. (\ref{con:eq5});
\IF {$c(\mathcal{T}(X, \tilde{X}_t, \alpha_{t, i}, -\beta_{t+1, i}, \mathcal{S}_t))=c(X)$}
\STATE $\bar{\beta}_{t, i+1}=\beta_{t, i+1}, \beta_{t, i+1}=\beta_{t, i}$;
\ENDIF
\ENDIF
\STATE $q=q+1$;
\STATE Update $\alpha_{t, i+1}$ based on Eq. (\ref{con:eq5});
\ENDFOR
\STATE $\tilde{X}_{t+1}=\mathcal{T}(X, \tilde{X}_t, \alpha_{t,i+1}, \beta_{t,i+1}, \mathcal{S}_t)$, $t=t+1$;
\ENDWHILE
\STATE \textbf{return} $\tilde{x}_t=\mathrm{IDCT}(\tilde{X}_t)$.
\end{algorithmic}
\end{algorithm}
To ensure successful initialization, we employ a targeted attack strategy by selecting the target example $x_t$ from a category different from that of the input image $x$, that is $c(x_t)\neq c(x)$. This target example will serve as our initial adversarial example $\tilde{x}_0=x_t$. Then, we apply the DCT to both the input image and the initial adversarial example:
\begin{equation}
X=\mathrm{DCT}(x),\quad \tilde{X}_0=\mathrm{DCT}(\tilde{x}_0),
\end{equation}
where $X$ is the frequency-domain representation of $x$ obtained through the DCT module.
To explore the distribution of gradient estimation between real and generated images, we utilize different frequency-domain segments as query subspaces for investigating the decision boundary based on the geometric information.

\subsubsection{Subspace Sampling}
We randomly sample a d-dimensional directional line $v_{\mathrm{rand}}$ in the frequency domain, which together with the current adversarial example $\tilde{X}_t$ direction determines a 2-D subspace $\mathcal{S}_t=\operatorname{span}(v_{\mathrm{rand}},\tilde{X}_t-X)$, where $\operatorname{span}$ is the vector space formed by all possible linear combinations of a given set of vectors. The core methodology is to employ triangular geometric properties for adversarial perturbation optimization. At each iteration, the input example $X$, current adversarial example $\tilde{X}_t$ and next adversarial example $\tilde{X}_{t+1}$ can naturally form a triangle in a subspace. According to the law of sines (suppose $a$, $b$ and $c$ are the side lengths of a triangle, and $\alpha$, $\beta$ and $\gamma$ are the opposite angles, we have $\frac{a}{\sin \alpha}=\frac{b}{\sin \beta}=\frac{c}{\sin \gamma}$), we have the following relationship:
\begin{equation}
\frac{\delta_t}{\sin \alpha_t}=\frac{\delta_{t+1}}{\sin \left(\pi-\left(\alpha_t+\beta_t\right)\right)},
\end{equation}
where $\delta_t=\|\tilde{X}_t-X\|_p$ is the current perturbation magnitude, $\alpha_t$ is the learned angle, $\beta_t$ is the search angle. To greedily decrease the perturbation(i.e., $\delta_{t+1}<\delta_t$), the following condition should satisfy the formulation as follows:
\begin{equation}
\beta_t+2 \alpha_t>\pi.
\end{equation}

\subsubsection{Candidate Triangle Search}
In the current subspace, we search for adversarial examples based on angle constraints. We first initialize the search angle $\beta_{t, 0}=\max (\pi-2 \alpha_t, \underline{\beta})$, where $\underline{\beta}=\pi/16$. If neither $\mathcal{T}(X, \tilde{X}_t, \alpha_t, \beta_{t,0}, \mathcal{S}_t)$ nor $\mathcal{T}(X, \tilde{X}_t, \alpha_t, -\beta_{t,0}, \mathcal{S}_t)$ is adversarial, we give up this subspace because it cannot bring any benefit, where $\mathcal{T}(\cdot)$ is the geometric transformation function that generates the next adversarial example based on triangle constraints in subspace $\mathcal{S}_t$. Otherwise, we use binary search to find the optimal angle $\beta^*$ in the interval $[\max (\pi-2 \alpha_t, \underline{\beta}), \min (\pi-\alpha_t, \pi / 2)]$. Finally, we calculate the position of the new adversarial example $\tilde{X}_{t+1}=\mathcal{T}(X, \tilde{X}_t, \alpha_t, \beta^*, \mathcal{S}_t)$.

\subsubsection{Adaptive Angle Adjustment}
Intuitively, the angle $\alpha$ balances the magnitude of perturbation and the difficulty to find an adversarial example, we dynamically adjust the learned angle $\alpha$ based on search results:
\begin{equation}
\alpha_{t,i+1} = 
\begin{cases}
\min(\alpha_{t,i}+\gamma, \frac{\pi}{2}+\tau) & \text{if } c(\tilde{X}_{t,i+1}) \neq c(X) \\[2pt]
\max(\alpha_{t,i}-\lambda\gamma, \frac{\pi}{2}-\tau) & \text{otherwise}
\end{cases}
\label{con:eq5}
\end{equation}
, where $\tilde{X}_{t,i+1}=\mathcal{T}(X, \tilde{X}_t, \alpha_{t,i}, \beta_{t}^*, \mathcal{S}_t)$ is the adversarial example generated by $\alpha_{t,i}$, $\gamma$ is the change rate, $\lambda$ is a constant, and $\tau$ restricts the upper and lower bounds of $\alpha$. We adopt $\lambda<1$ to prevent decreasing the angle too fast considering much more failures than successes during the perturbation optimization. Note that the larger angle $\alpha$ makes it harder to find an adversarial example. However, a too small angle $\alpha$ results in a much lower bound for $\beta$, which also makes $\mathcal{T}(X, \tilde{X}_t, \alpha_t, \beta^*, \mathcal{S}_t)$ far away from the current adversarial example $\tilde{X}_t$, decreasing the probability to find an adversarial example. Thus, we add the bounds for $\alpha$ to restrict it in an appropriate range. All geometric operations are performed in DCT frequency domain, then mapped back to image space via Inverse Discrete Cosine Transform (IDCT):
\begin{equation}
\tilde{x}=\mathrm{IDCT}(\tilde{X}),
\end{equation}
where $\tilde{x}$ is the adversarial example that is finally transformed back to the spatial domain.

Our research is inspired by the different frequency domain characteristics of real images and generated images. After conducting hypothesis-informed experiments on different frequency domain subspaces, we find that the gradient estimation of real images is more concentrated in the low-frequency and high-frequency parts, while the gradient estimation of generated images is more concentrated in the low-frequency part. In view of this, we use the high-low mixed frequency-domain as the subspace for real images and the low frequency-domain subspace for generated images. We iteratively search for candidate triangles to find adversarial samples and update the angle $\alpha$ accordingly.

\subsection{Efficient Initialization Based on ``Adversarial Example Soup''}
To further improve query efficiency, we introduce the paradigm of adversarial example soup~\cite{10858076} for our efficient initialization. In AIGC-related security research, transfer-based attacks can generate adversarial samples via public surrogate models and directly launch black-box attacks against target detectors~\cite{diao2025vulnerabilitiesaigeneratedimagedetection}. But when the surrogate model and the target model differ significantly in structure, the attack effectiveness may be compromised. Our approach fundamentally differs from conventional transfer-based attacks by repurposing their attack outputs exclusively as initialization inputs for our framework. Initializing decision-based attacks with the results of transfer-based attacks can ensure both attack effectiveness and query efficiency.

We use the variable $\theta$ to represent the parameters of the surrogate model $F$. For optimization, $L(\theta, x, y)$ denotes the loss function of the surrogate model. The formulation of the generating adversarial examples $x^{\mathrm{adv}}$ can be written as follows:
\begin{equation}
\underset{x^{\mathrm{a d v}}}{\arg \max }\ L\left(\theta, x^{\mathrm{a d v}}, y\right), \quad \text { s.t. }\left\|x^{\mathrm{a d v}}-x\right\|_p \leq \varepsilon ,
\end{equation}
where $x$ is the original image, $y$ is the ground label, $\varepsilon$ is the allowed maximum perturbation.

During the generation of adversarial examples, we take the Momentum Iterative Gradient (MIG) method as an example~\cite{10378472}. At each iteration, MIG updates the adversarial example by accumulating momentum (exponentially weighted average of past gradients):
\begin{equation}
g_{t+1}=\mu \cdot g_t+\frac{\nabla_x L\left(\theta, x_t^{\mathrm{a d v}}, y\right)}{\left\|\nabla_x L\left(\theta, x_t^{\mathrm{a d v}}, y\right)\right\|_1},
\end{equation}

\begin{equation}
x_{t+1}^{\mathrm{a d v}}=\operatorname{Clip}_{x, \varepsilon}\left(x_t^{\mathrm{a d v}}+\eta \cdot \operatorname{sign}\left(g_{t+1}\right)\right),
\end{equation}
where $g_t$ denotes the momentum gradient at the t-th iteration, $\mu$ is the momentum decay factor, $\eta$ is the step size, $x_t^{\mathrm{a d v}}$ is the adversarial example at the t-th iteration. Then, our method preserves $n$ adversarial examples from both optimal and suboptimal iterations, generating the final adversarial example $x_{avg}^{\mathrm{adv}}$ through ensemble averaging. The formulation is $x_{avg}^{\mathrm{adv}}=\sum_{i=1}^n w_i x_i^{\mathrm{a d v}}$, where it satisfies $w_i=1/n$. In the final stage, the generated $x_{avg}^{\mathrm{adv}}$ is given to our detection model for evaluation. When the sample soup is adversarial, that is $c(x_{avg}^{\mathrm{adv}})\neq c(x)$, it serves as the initial point for the decision-based attack. In cases of failure, the algorithm defaults back to the standard targeted attack procedure. Our proposed FBA$^2$D method is presented as Algorithm \ref{alg:algorithm}. Using triangular geometry properties and frequency domain optimization, it exposes vulnerabilities in AIGC detectors. Concurrently, we accelerate attacks using adversarial example soups or transfer-based attacks. 

\section{Experiment}
\subsection{Experiment Settings}
\subsubsection{Datasets}
We choose Synthetic LSUN dataset~\cite{Wang_2020_CVPR} and GenImage dataset~\cite{zhu2023genimage} as our evaluation benchmarks. Details of the datasets are provided in the appendix.
\subsubsection{Evaluated Models}
We evaluate our proposed method on the following prominent AIGC detector architectures: CNNSpot~\cite{Wang_2020_CVPR}, DenseNet~\cite{huang2017densely}, EfficientNet~\cite{tan2019efficientnet}, MobileNet~\cite{howard2017mobilenets}, Vision Transformer (ViT)~\cite{dosovitskiy2021image}, Swin Transformer~\cite{liu2021swin}, PatchCraft~\cite{zhong2024patchcraftexploringtexturepatch}, AIDE~\cite{DBLP:conf/iclr/YanLCHJ0X25}, and Effort~\cite{yan2025orthogonalsubspacedecompositiongeneralizable}.
\subsubsection{Compared Methods}
We employ the HopSkipJumpAttack (HSJA) method~\cite{chen2020hopskipjumpattack}, which leverages high-dimensional geometric properties. Two frequency-based attack methods are Geometric-based Attack~\cite{rahmati2020geoda} and TA~\cite{10.1007/978-3-031-20065-6_10} methods. OPT is a milestone generic and optimization-based hard-label black box attack method~\cite{cheng2018query}. Sign-OPT~\cite{cheng2020sign} resolves the query efficiency bottleneck in hard-label black-box attacks through its core principle of gradient sign estimation coupled with directional aggregation. We also include the Approximation Decision Boundary Attack (ADBA) method~\cite{wang2025adba}, which uses the distribution’s median value as the approximation to differentiate the perturbation directions with high query efficiency. To ensure a fair comparison, all attacks are limited to 500 queries, and we report the attack success rate, the average number of queries, and the average $\ell_{2}$ distance at the Root Mean Squared Error (RMSE) threshold of 0.1, 0.05, and 0.01. The implementation details of our experiment are provided in the appendix.

\begin{table*}[!t]
  \centering
  \setlength{\tabcolsep}{2.7pt}  
  \renewcommand{\arraystretch}{1.1} 

  \begin{tabular}{c|c|*{15}{c}}  
    \toprule 
    \multirow{2}{*}{Dataset}   & Model
      & \multicolumn{3}{c}{CNNSpot}
      & \multicolumn{3}{c}{MobileNet}
      & \multicolumn{3}{c}{DenseNet}
      & \multicolumn{3}{c}{ViT}
      & \multicolumn{3}{c}{EfficientNet}\\
    \cmidrule(lr){3-5}\cmidrule(lr){6-8}\cmidrule(lr){9-11}
    \cmidrule(lr){12-14}\cmidrule(lr){15-17}
    & RMSE
      & 0.1   & 0.05  & 0.01
      & 0.1   & 0.05  & 0.01
      & 0.1   & 0.05  & 0.01
      & 0.1   & 0.05  & 0.01
      & 0.1   & 0.05  & 0.01\\
    \midrule
    \multirow{7}{*}{LSUN}   
    & OPT
      & 0.768 & 0.522 & 0.006
      & 0.836 & 0.587 & 0.037
      & 0.666 & 0.503 & 0.006
      & 0.748 & 0.323 & 0.001
      & 0.703 & 0.525 & 0.022 \\
    & Sign-OPT
      & 0.170 & 0.020 & 0.000
      & 0.296 & 0.078 & 0.004
      & 0.176 & 0.014 & 0.000
      & 0.259 & 0.043 & 0.001
      & 0.226 & 0.044 & 0.003 \\
    & HSJA
      & 0.809 & 0.509 & 0.004
      & 0.541 & 0.419 & 0.054
      & 0.497 & 0.449 & 0.004
      & 0.958 & 0.660 & 0.074
      & 0.527 & 0.341 & 0.021 \\
    & GeoDA
      & 0.521 & 0.521 & 0.346
      & 0.547 & 0.547 & 0.485
      & 0.517 & 0.517 & 0.389
      & 0.581 & 0.581 & 0.473
      & 0.647 & 0.552 & 0.349 \\
    & TA
      & 0.766 & 0.673 & 0.310
      & 0.931 & 0.889 & 0.365
      & 0.558 & 0.551 & 0.417
      & 0.990 & 0.985 & 0.731
      & 0.656 & 0.558 & 0.184 \\
     & ADBA
      & 0.498 & 0.494 & 0.000
      & 0.492 & 0.477 & 0.000
      & 0.498 & 0.489 & 0.000
      & 0.502 & 0.497 & 0.000
      & 0.489 & 0.475 & 0.000 \\
    & FBA$^2$D
      & \textbf{0.979} & \textbf{0.973} & \textbf{0.909}
      & \textbf{0.986} & \textbf{0.972} & \textbf{0.796}
      & \textbf{0.959} & \textbf{0.954} & \textbf{0.874}
      & \textbf{0.999} & \textbf{0.996} & \textbf{0.777}
      & \textbf{0.770} & \textbf{0.718} & \textbf{0.491} \\
    \midrule
    \multirow{6}{*}{GenImage}   
    & Sign-OPT
      & 0.281 & 0.087 & 0.002
      & 0.076 & 0.015 & 0.002
      & 0.231 & 0.052 & 0.001
      & 0.258 & 0.115 & 0.021
      & 0.395 & 0.113 & 0.001 \\
    & HSJA
      & 0.520 & 0.410 & 0.063
      & 0.532 & 0.324 & 0.049
      & 0.488 & 0.337 & 0.043
      & 0.429 & 0.275 & 0.071
      & 0.605 & 0.490 & 0.050 \\
    & GeoDA
      & 0.658 & 0.479 & 0.187
      & 0.803 & 0.715 & 0.370
      & 0.555 & 0.377 & 0.219
      & 0.527 & 0.488 & 0.239
      & 0.668 & 0.536 & 0.195 \\
    & TA
      & 0.902 & 0.692 & 0.221
      & 0.826 & 0.650 & 0.216
      & 0.858 & 0.723 & 0.266
      & \textbf{0.933} & \textbf{0.761} & \textbf{0.367}
      & 0.822 & 0.567 & 0.113 \\
    & ADBA
      & 0.698 & 0.504 & 0.000
      & 0.795 & 0.643 & 0.000
      & 0.427 & 0.336 & 0.000
      & 0.490 & 0.473 & 0.000
      & 0.617 & 0.488 & 0.000\\
    & FBA$^2$D
      & \textbf{0.908} & \textbf{0.695} & \textbf{0.314}
      & \textbf{0.962} & \textbf{0.882} & \textbf{0.566}
      & \textbf{0.999} & \textbf{0.954} & \textbf{0.674}
      & 0.905 & 0.740 & 0.314
      & \textbf{0.923} & \textbf{0.847} & \textbf{0.368} \\
    \bottomrule
  \end{tabular}

  \caption{Attack Success Rate (ASR) of Different Black-box Attack Methods Across Multiple Models and Datasets part \uppercase\expandafter{\romannumeral 1}.}
  \label{tab:attack_comparison1}
\end{table*}

\begin{table*}[!t]
  \centering
  \setlength{\tabcolsep}{5.2pt}  
  \renewcommand{\arraystretch}{1.1} 

  \begin{tabular}{c|c|*{12}{c}}  
    \toprule 
    \multirow{2}{*}{Dataset} & Model
      & \multicolumn{3}{c}{Swin Transformer}
      & \multicolumn{3}{c}{PatchCraft}
      & \multicolumn{3}{c}{AIDE}
      & \multicolumn{3}{c}{Effort} \\
    \cmidrule(lr){3-5}\cmidrule(lr){6-8}\cmidrule(lr){9-11}
    \cmidrule(lr){12-14}
    & RMSE
      & 0.1 & 0.05 & 0.01
      & 0.1 & 0.05 & 0.01
      & 0.1 & 0.05 & 0.01
      & 0.1 & 0.05 & 0.01 \\
    \midrule
    \multirow{7}{*}{LSUN}
    & OPT
      & 0.817 & 0.535 & 0.002
      & 0.792 & 0.518 & 0.003
      & 0.805 & 0.526 & 0.002
      & 0.788 & 0.512 & 0.004 \\
    & Sign-OPT
      & 0.344 & 0.028 & 0.000
      & 0.318 & 0.025 & 0.000
      & 0.329 & 0.026 & 0.000
      & 0.335 & 0.027 & 0.000 \\
    & HSJA
      & 0.676 & 0.624 & 0.019
      & 0.692 & 0.608 & 0.021
      & 0.684 & 0.616 & 0.020
      & 0.688 & 0.612 & 0.022 \\
    & GeoDA
      & 0.522 & 0.522 & 0.500
      & 0.515 & 0.515 & 0.492
      & 0.518 & 0.518 & 0.496
      & 0.520 & 0.520 & 0.498 \\
    & TA
      & 0.834 & 0.778 & 0.299
      & 0.821 & 0.765 & 0.285
      & 0.827 & 0.771 & 0.292
      & 0.830 & 0.774 & 0.295 \\
    & ADBA
      & 0.483 & 0.473 & 0.000
      & 0.491 & 0.481 & 0.000
      & 0.487 & 0.477 & 0.000
      & 0.485 & 0.475 & 0.000 \\
    & FBA$^2$D
      & \textbf{0.900} & \textbf{0.890} & \textbf{0.755}
      & \textbf{0.885} & \textbf{0.875} & \textbf{0.742}
      & \textbf{0.892} & \textbf{0.882} & \textbf{0.748}
      & \textbf{0.895} & \textbf{0.885} & \textbf{0.751} \\
    \midrule
    \multirow{6}{*}{GenImage}
    & Sign-OPT
      & 0.355 & 0.149 & 0.002
      & 0.342 & 0.138 & 0.001
      & 0.348 & 0.143 & 0.002
      & 0.351 & 0.146 & 0.002 \\
    & HSJA
      & \textbf{0.992} & 0.839 & 0.264
      & 0.978 & 0.825 & 0.252
      & 0.985 & 0.832 & 0.258
      & 0.988 & 0.835 & 0.261 \\
    & GeoDA
      & 0.801 & 0.784 & 0.343
      & 0.788 & 0.771 & 0.331
      & 0.794 & 0.777 & 0.337
      & 0.797 & 0.780 & 0.340 \\
    & TA
      & 0.547 & 0.443 & 0.129
      & 0.534 & 0.430 & 0.117
      & 0.540 & 0.436 & 0.123
      & 0.543 & 0.439 & 0.126 \\
    & ADBA
      & 0.649 & 0.507 & 0.000
      & 0.636 & 0.494 & 0.000
      & 0.642 & 0.500 & 0.000
      & 0.645 & 0.503 & 0.000 \\
    & FBA$^2$D
      & 0.987 & \textbf{0.981} & \textbf{0.891}
      & \textbf{0.995} & \textbf{0.974} & \textbf{0.868}
      & \textbf{0.991} & \textbf{0.977} & \textbf{0.879}
      & \textbf{0.989} & \textbf{0.979} & \textbf{0.885} \\
    \bottomrule
  \end{tabular}

  \caption{Attack Success Rate (ASR) of Different Black-box Attack Methods Across Multiple Models and Datasets part \uppercase\expandafter{\romannumeral 2}.}
  \label{tab:attack_comparison2}
\end{table*}

\begin{table*}[!t]
    \centering
    \setlength{\tabcolsep}{5.6pt}  
    \begin{tabular}{c*{9}{c}}
    \toprule
    Model       
    & \multicolumn{3}{c}{CNNSpot}        
    & \multicolumn{3}{c}{MobileNet}
    & \multicolumn{3}{c}{ViT} \\ 
    \cmidrule(lr){2-4}\cmidrule(lr){5-7}\cmidrule(lr){8-10}
    RMSE        
    & 0.1       & 0.05      & 0.01
    & 0.1       & 0.05      & 0.01
    & 0.1       & 0.05      & 0.01       \\ \hline
    10\%L       
    & 0.53/1.00 & 0.36/0.99 & 0.13/0.49 
    & 0.86/1.00 & 0.81/0.97 & \textbf{0.38}/0.35
    & 0.98/1.00 & 0.97/1.00 & 0.92/0.53  \\
    20\%L       
    & 0.49/1.00 & 0.18/1.00 & 0.06/\textbf{0.69} 
    & 0.61/0.98 & 0.47/0.98 & 0.28/0.82
    & 0.64/1.00 & 0.58/1.00 & 0.51/\textbf{0.84}  \\
    30\%L       
    & 0.29/1.00 & 0.07/1.00 & 0.02/0.61 
    & 0.14/0.99 & 0.09/0.99 & 0.04/\textbf{0.88}
    & 0.47/1.00 & 0.42/1.00 & 0.32/0.81  \\
    10\%H       
    & 0.17/0.55 & 0.03/0.14 & 0.00/0.03 
    & 0.53/0.61 & 0.23/0.17 & 0.01/0.02
    & 0.62/0.63 & 0.20/0.16 & 0.03/0.02  \\
    80\%M       
    & 0.35/1.00 & 0.17/1.00 & 0.07/0.28  
    & 0.85/0.99 & 0.71/0.99 & 0.35/0.72
    & 0.91/1.00 & 0.72/1.00 & 0.27/0.30  \\
    10\%L+5\%H  
    & 0.69/1.00 & 0.52/1.00 & \textbf{0.19}/0.48  
    & 0.90/0.98 & 0.85/0.95 & 0.32/0.30
    & \textbf{0.99}/1.00 & \textbf{0.99}/1.00 & \textbf{0.93}/0.50  \\
    10\%L+10\%H 
    & \textbf{0.70}/1.00 & \textbf{0.54}/1.00 & 0.15/0.42 
    & \textbf{0.95}/1.00 & \textbf{0.90}/0.94 & 0.28/0.27
    & \textbf{0.99}/1.00 & \textbf{0.99}/0.98 & 0.89/0.44  \\
    10\%L+15\%H 
    & 0.68/1.00 & 0.52/1.00 & 0.12/0.38
    & 0.91/0.98 & 0.85/0.91 & 0.23/0.23
    & 0.98/1.00 & 0.98/0.97 & 0.85/0.40  \\ 
    \bottomrule
    \end{tabular} 
    \caption{ASR of different frequency element combinations.}  
    \label{tab:frequency_comparison}
\end{table*}

\subsection{Experimental Results}
\par The experiments we would conduct are primarily designed to address several questions concerning decision-based attacks on AIGC detectors. Adversarial attacks against AIGC detectors exhibit markedly different characteristics from those observed in conventional image classification. AIGC detectors must grapple with more complex multimodal data and generator-specific attributes, whereas image classifiers primarily focus on pixel-level perturbations and gradient estimation. This divergence stems from two key factors. First, there exists an inherent asymmetry in the detection task. The real world contains a vast number of real images, whereas synthetic images are relatively scarce. Second, image classifiers depend on semantic understanding to recognize objects. Although AI-generated images share similar semantic features with real images, they evoke different perceptual experiences. Prior work has attributed this discrepancy to the relative scarcity of high-frequency components in generated images~\cite{pmlr-v119-frank20a,Chandrasegaran_2021_CVPR}. Building on these insights, our experiments will systematically investigate the following questions.
\begin{itemize}
\item \textbf{Q}: Considering the fundamental differences between AIGC detection and conventional classification, and the divergent frequency domain signatures of generated versus real images, how can decision-based attacks with frequency domain transformation be more effectively tailored to the AIGC detection task? \textbf{A}: We conduct experiments across multiple frequency bands to analyze how the signal information of generated and real images is distributed throughout the frequency spectrum.
\item \textbf{Q}: How do we evaluate the effectiveness of the method? \textbf{A}: We perform a comprehensive, multi-dimensional assessment, considering the attack success rate, the number of queries, and the visual quality of the adversarial examples.
\item \textbf{Q}: What is the purpose of the adversarial example soup with the initialization of transfer-based attack? \textbf{A}: We introduce the concepts of adversarial example soups and transfer-based attacks to optimize the initialization of decision-based attacks, thereby improving query efficiency.
\end{itemize}

\subsubsection{Main Results}
\par Table \ref{tab:attack_comparison1} and Table \ref{tab:attack_comparison2} show the attack results of the mainstream decision-based attack methods on the LSUN~\cite{Wang_2020_CVPR} and GenImage~\cite{zhu2023genimage} benchmarks. On the GenImage dataset, we train the model using the entire dataset, and we conduct the evaluation on the SDv1.4 subset (images generated by the StableDiffusion model~\cite{rombach2022high}). 
The evaluated methods are HSJA~\cite{chen2020hopskipjumpattack}, GeoDA~\cite{rahmati2020geoda}, TA~\cite{10.1007/978-3-031-20065-6_10}, ADBA~\cite{wang2025adba}, OPT~\cite{cheng2018query}, Sign-OPT~\cite{cheng2020sign}, and our proposed method. The attack models are CNNSpot~\cite{Wang_2020_CVPR}, DenseNet~\cite{huang2017densely}, EfficientNet~\cite{tan2019efficientnet}, MobileNet~\cite{howard2017mobilenets}, ViT~\cite{dosovitskiy2021image}, Swin Transformer~\cite{liu2021swin}, PatchCraft~\cite{zhong2024patchcraftexploringtexturepatch}, AIDE~\cite{DBLP:conf/iclr/YanLCHJ0X25} and Effort~\cite{yan2025orthogonalsubspacedecompositiongeneralizable}. 
The testing RMSEs are 0.1, 0.05, and 0.01. Our proposed method achieves the optimal attack results. It validates the effectiveness of the integration of adversarial example soups and decision-based attacks based on the frequency-domain signals. The attack effectiveness of our proposed method is more evident on the LSUN dataset~\cite{Wang_2020_CVPR} than on the GenImage dataset~\cite{zhu2023genimage}. The potential reason is that the samples of the LSUN dataset are generated by the ProGAN model~\cite{karras2018progressive}, and the samples of the GenImage dataset are crafted by the StableDiffusion model~\cite{rombach2022high} with more powerful generative abilities. Our method also reveals the adversarial vulnerability of three state-of-the-art AIGC detection models, PatchCraft~\cite{zhong2024patchcraftexploringtexturepatch}, AIDE~\cite{DBLP:conf/iclr/YanLCHJ0X25}, and Effort~\cite{yan2025orthogonalsubspacedecompositiongeneralizable}. 
\par
\begin{figure}[!t]
    \centering
    \includegraphics[width=1\linewidth]{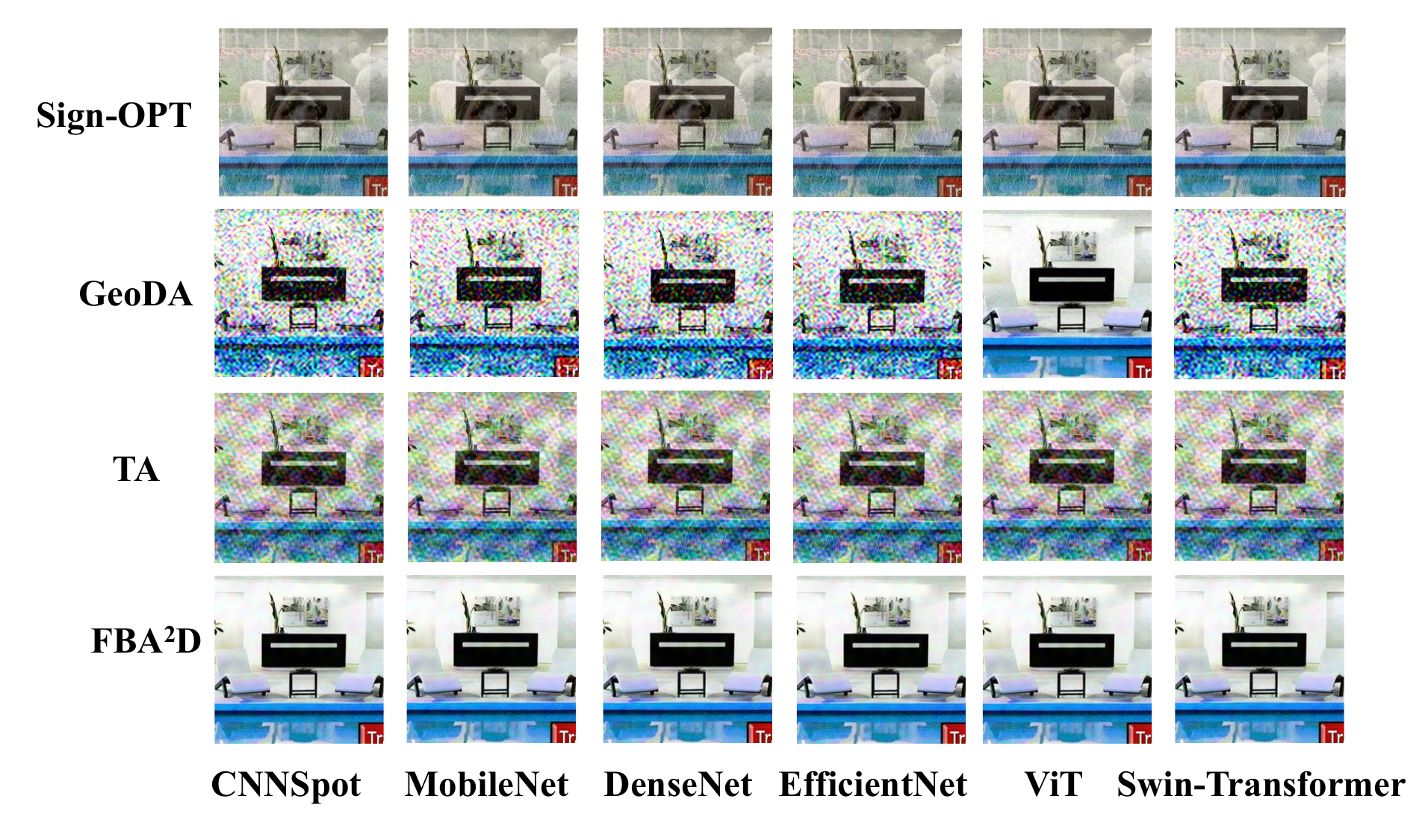}
    \caption{The visualization results of adversarial examples with different method.}
    \label{fig:placeholder}
\end{figure}

 Figure \ref{fig:placeholder} shows the adversarial example visualization results between our method and other baseline methods, including two frequency-domain-based methods: GeoDA~\cite{rahmati2020geoda} and TA~\cite{10.1007/978-3-031-20065-6_10}. The noise produced by GeoDA~\cite{rahmati2020geoda} is highly conspicuous, filling the image with random speckles. In contrast, Sign-OPT~\cite{cheng2020sign}, TA~\cite{10.1007/978-3-031-20065-6_10}, and FBA$^{2}$D introduce much subtler perturbations. Especially, the experimental results of FBA$^{2}$D are nearly indistinguishable from the original image. It can be clearly seen that The adversarial examples generated by our proposed method have better concealment.
\subsubsection{Frequency Perspectives}
Table \ref{tab:frequency_comparison} illustrates the experimental results related to the frequency perspective. The experimental results indicate that, for attacks on real images, selecting 10\% of the high-frequency and 10\% of the low-frequency components yields the best performance. For attacks on generated images, selecting 20\% of the low-frequency components achieves the highest effectiveness.
\subsubsection{Ablation Study}
\begin{figure}
    \centering
    \includegraphics[width=1\linewidth]{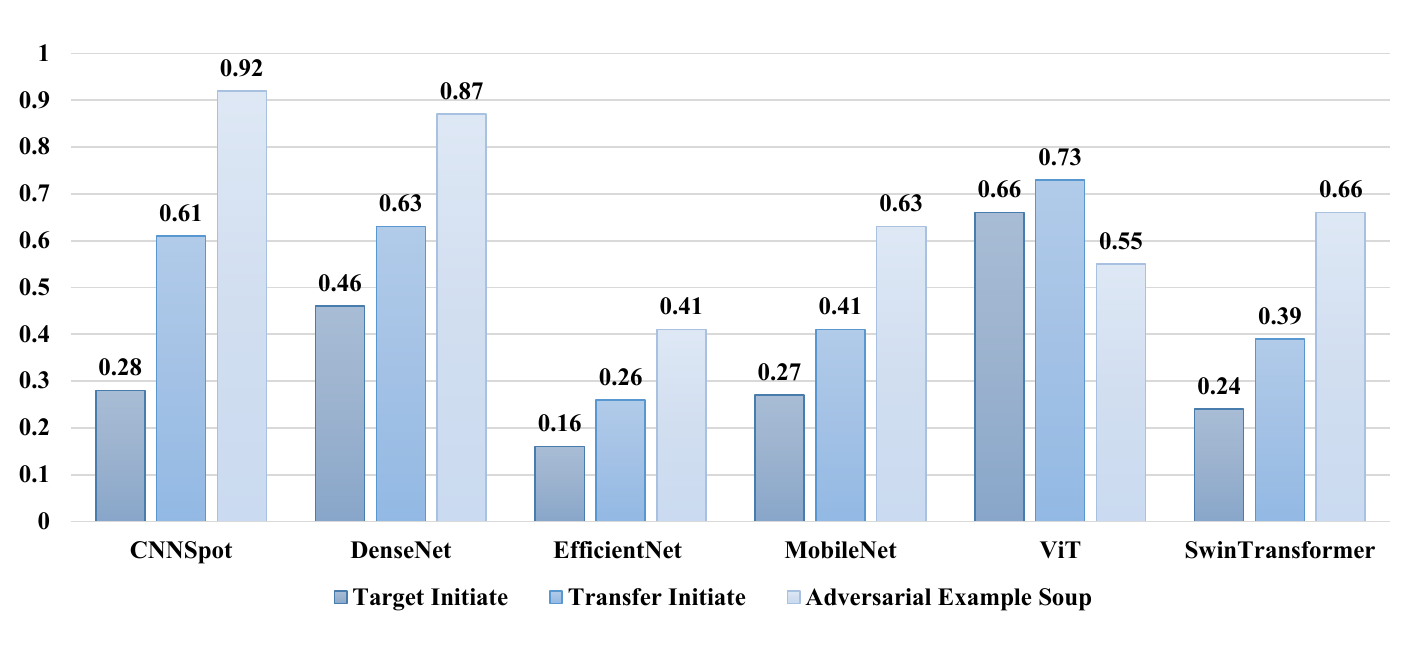}
    \caption{Ablation Study on Different Initialization Methods.}
    \label{fig:ablation_study}
\end{figure}

Fig.\ref{fig:ablation_study} shows the ablation study result of the diverse attack modules. It shows the impact of different initialization methods on attack success rate. For most models, initialization using the sample soup method yields the best results, while initialization using the target attack method yields the worst results. When the RMSE is set to 0.05, the attack configuration of adversarial example soup woks is set on most of the target model except in ViT~\cite{dosovitskiy2021image}. 
The potential reason is ViT's architectural divergence from GramNet~\cite{liu2020globaltextureenhancementfake}, the source of the adversarial example soup.

\subsection{Discussion}
\par The experimental results show that the frequency-domain dependence of the generated image and the real image is indeed different, which further verifies the basis of our method design. We also find that the decision-based attack method originally used for classification tasks selects the low-frequency (the top 10\% low-frequency) as the query subspace, which is an empirically optimal choice. 
In AIGC detection tasks, there exists a notable distribution discrepancy in frequency-domain signal information between real and generated images~\cite{pmlr-v119-frank20a,Chandrasegaran_2021_CVPR}. When handling real images, the AIGC detection model predominantly focuses on mixed high-low frequency domain information. In contrast, to detect generative images, the model relies mainly on low-frequency components (specifically, the highest 20\% lowest frequencies). Although our proposed decision‐based black‐box attack method has exposed security issues in AIGC applications, current methods still exhibit certain limitations. First, although most adversarial example soup attacks have proven effective, there still exists an ``overfitting'' phenomenon on ViT and EfficientNet. Second, our frequency‐domain signal processing remains empirical and manual, lacking theoretical explanations. Finally, attacks on diffusion models could be more effective.
\section{Conclusion}
\par This paper proposes an innovative decision-based attack method for the AIGC detector. Our proposed FBA$^2$D method is guided by the signal decomposition method in the frequency domain. The ensemble of adversarial example soups and initialization with transfer-based attack can improve the efficiency of decision-based attacks. Our evaluation of the Synthetic LSUN dataset and GenImage dataset reveals the security risks faced by the new media industry based on AIGC. Quantitative and qualitative experiments have shown that our proposed method outperforms other attack methods in terms of attack effectiveness and concealment of generated adversarial samples, especially other frequency-domain-based attack methods.
\par However, our study has certain limitations. First, the transfer-based attacks using the adversarial example soups are not effective against target models whose architectures differ substantially from the source model (e.g., CNN$\rightarrow$ViT). Second, although we use CNNSpot to investigate frequency-domain signal characteristics, other models in real-world scenarios exhibit different frequency components compared with CNNSpot.
\par In future work, we would improve the effectiveness of decision-based attacks against AIGC detectors with diverse architectures. At the same time, exploring paradigms for effective defense is also worthwhile. Investigating the differences between generated and real images in both low- and high-frequency domains and their impact on adversarial robustness warrants further theoretical and empirical study.
\clearpage
\setcounter{page}{1}
\maketitlesupplementary

%
In this appendix, we provide more study details relevant to our proposed FBA$^2$D method. The appendix introduces additional experimental settings, and experimental results.
\begin{itemize}
\item \textbf{Experimental Datasets Additional Information} -- Detailed composition of the datasets.
\item \textbf{Experimental Settings Additional Information} -- Description of our choices of experimental settings.
\item \textbf{Experimental Results Additional Information} –- An extended analysis of the experimental results from the main manuscript.
\end{itemize}

\section{Datasets}
The Synthetic LSUN dataset contains 720k real images from LSUN and 360k fake images generated by ProGAN~\cite{karras2018progressive}. GenImage~\cite{NEURIPS2023_f4d4a021} comprises more than one million pairs of AI-generated fake images and their corresponding real images, spanning all 1,000 classes of ImageNet~\cite{deng2009imagenet}. These two datasets are crucial benchmarks for understanding the robustness of AIGC.

\section{Implementation Details}
In our decision-based black-box attack scenarios, we set the momentum decay to 0.95, run for 10 iterations, use a maximum perturbation of $\varepsilon=8.0/255.0$ and a scaling factor 
$S_{F}=20$. For transfer-based attacks with adversarial example soups, we build the soup by averaging adversarial examples generated by Momentum Integrated Gradients (MIG)~\cite{10378472} on the substitute model GramNet~\cite{liu2020globaltextureenhancementfake} after 6, 7, 8, 9, and 10 iterations. If the adversarial example soup initialization succeeds, we launch the decision-based attack from the clean image using the soup. All hyperparameters for HSJA~\cite{chen2020hopskipjumpattack}, GeoDA~\cite{rahmati2020geoda}, OPT~\cite{cheng2018query}, Sign-OPT~\cite{cheng2020sign}, and ADBA~\cite{wang2025adba} follow their original study settings. The experiments were run on four NVIDIA GeForce RTX 3090 GPUs.

\section{Comparison of Frequency Domain Transformation Methods}
In this section, we provide experimental results under different frequency-domain transformation methods and analyze the impact of frequency-domain transformation methods on the experimental results. In our experiments, we use two frequency-domain transformation methods. They are Discrete Cosine Transform (DCT), and Discrete Fourier Transform (DFT). We explore different frequency domain transformation methods to assess their impact on the quality of generated adversarial examples. For consistency, we maintain the same experimental procedure. Table \ref{tab:study} illustrates the result that with the 10\% L, 20\% L, and 10\% L + 10\% H configurations, the DCT-based method consistently outperforms the DFT-based approach.

\begin{table*}[!]
    \centering
    \caption{Across Multiple Frequency Subspaces Using Different Frequency-domain Transformation Methods on Real (left) and Fake Images (right)}
    \begin{tabular}{ccccccc}
    \hline
    Method & \multicolumn{3}{c}{DCT}  & \multicolumn{3}{c}{DFT}                \\ \cmidrule(lr){2-4}\cmidrule(lr){5-7}
    RMSE   & 0.1   & 0.05   & 0.01
           & 0.1   & 0.05   & 0.01        \\ \hline
    10\% L
    & \textbf{0.533} / \textbf{1.000}     & \textbf{0.356} / \textbf{0.990}    & \textbf{0.130} / \textbf{0.490} 
    & 0.348 / 0.984     & 0.090 / 0.984    & 0.033 / 0.397\\
    10\% L + 10\% H
    & \textbf{0.697} / \textbf{1.000}     & \textbf{0.540} / 0.990    & \textbf{0.148} / \textbf{0.419}
    & 0.517 / 0.995     & 0.382 / \textbf{0.995}    & 0.106 / 0.330\\
    20\% L
    & \textbf{0.490} / \textbf{1.000}     & \textbf{0.177} / \textbf{1.000}    & \textbf{0.057} / \textbf{0.687}
    & 0.268 / 0.990     & 0.074 / 0.990    & 0.024 / 0.351\\ 
    \hline
    
    \label{tab:study}
    \end{tabular}
\end{table*}

\begin{figure}[!]
    \centering
\includegraphics[width=1\linewidth]{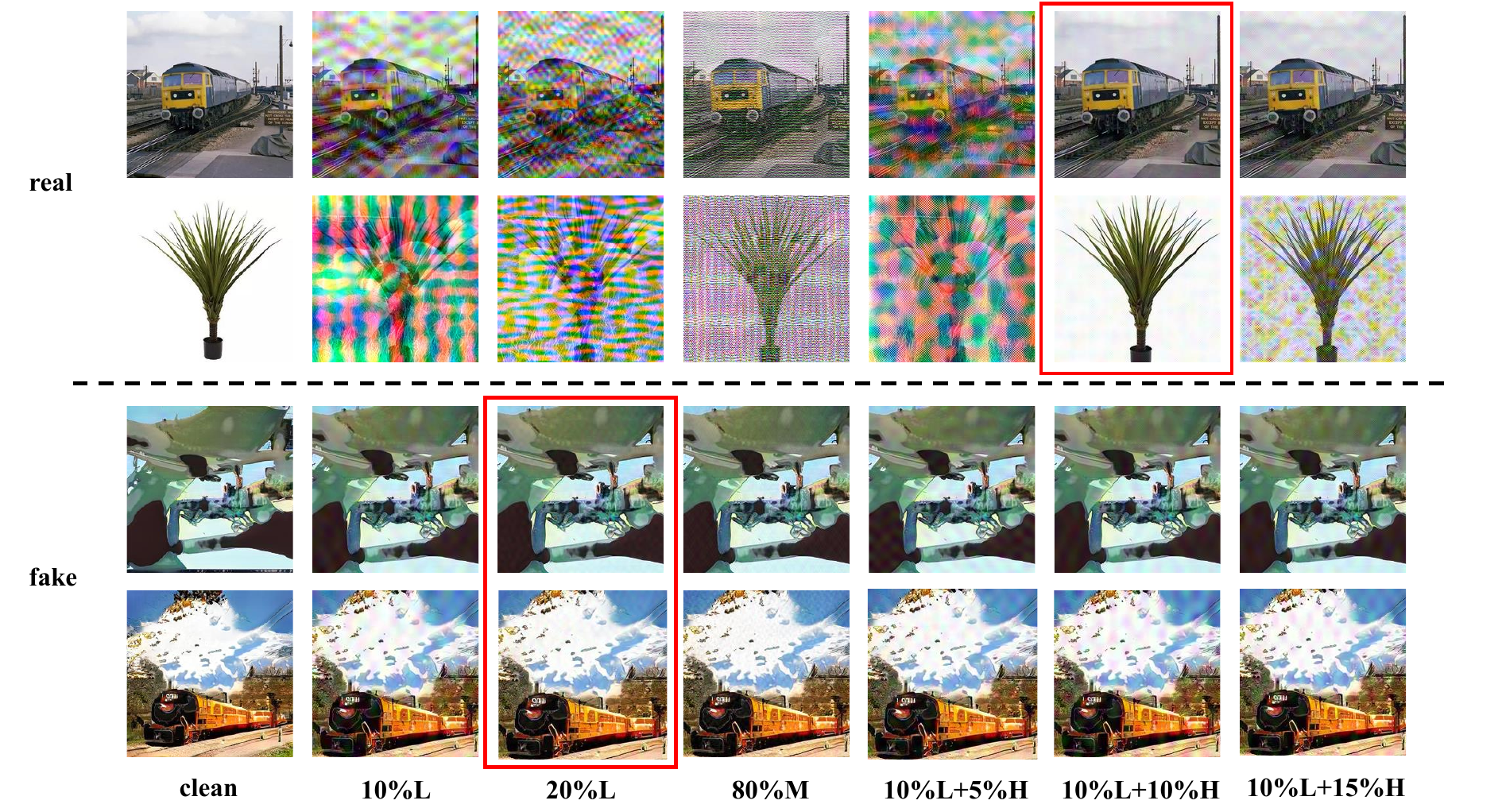}
    \caption{The visualization results of adversarial examples with different frequency element combinations of CNNSpot.}
    \label{fig:CNNSpot}
\end{figure}

\begin{figure}
    \centering
    \includegraphics[width=1\linewidth]{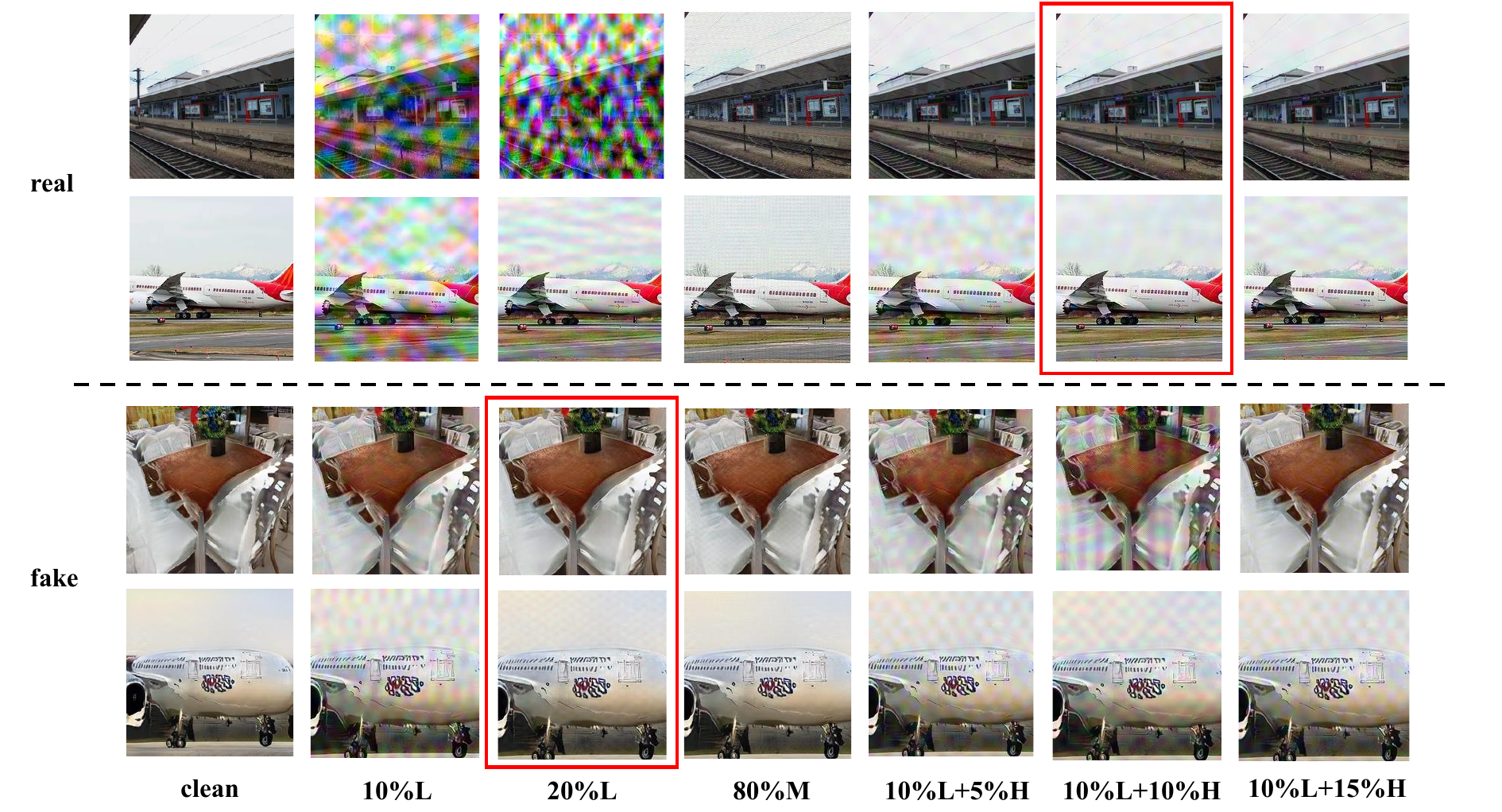}
    \caption{The visualization results of adversarial examples with different frequency element combinations of MobileNet.}
    \label{fig:Mobilenet}
\end{figure}

\begin{table*}[htbp]
  \centering
  \small
  \caption{Transfer-based Attack Results after Sample-Soup Augmentation and Clean-Sample Replacement (Average over Iterations 6–10 and 8–12)}
  \label{tab:transfer-attacks}
  \begin{tabular}{l*{6}{c}}
    \toprule
    \multirow{3}{*}{Iterations} 
    & \multicolumn{6}{c}{Target Models \& Metrics (ASR / Queries / L2)} \\
    \cmidrule(lr){2-7}
    & \multicolumn{1}{c}{CNNSpot} 
    & \multicolumn{1}{c}{DenseNet} 
    & \multicolumn{1}{c}{EfficientNet} 
    & \multicolumn{1}{c}{MobileNet} 
    & \multicolumn{1}{c}{ViT} 
    & \multicolumn{1}{c}{Swin Transformer} \\

    \midrule
    
    \multirow{3}{*}{\centering 6--10} 
    & 0.994 / 512 / 2.66 & 0.966 / 526 / 2.51 & 0.813 / 701 / 5.20 & 0.993 / 638 / 4.15 & 0.999 / 724 / 4.24 & 0.910 / 574 / 3.44 \\
    & 0.991 / 511 / 2.58 & 0.965 / 525 / 2.48 & 0.789 / 692 / 4.47 & 0.977 / 631 / 3.84 & 0.991 / 722 / 4.09 & 0.908 / 573 / 3.37 \\
    & 0.919 / 507 / 2.43 & 0.870 / 508 / 2.23 & 0.414 / 506 / 2.62 & 0.630 / 513 / 2.48 & 0.552 / 504 / 1.49 & 0.660 / 501 / 2.70 \\
    
    \addlinespace[0.5em]
    
    \multirow{3}{*}{\centering 8--12} 
    & 0.994 / 511 / 2.64 & 0.966 / 526 / 2.51 & 0.810 / 701 / 5.20 & 0.993 / 637 / 4.15 & 0.998 / 728 / 4.29 & 0.910 / 574 / 3.44 \\
    & \textbf{0.992} / 510 / 2.59 & 0.965 / 525 / 2.48 & 0.789 / 692 / 4.47 & 0.977 / 631 / 3.84 & 0.990 / 726 / 4.16 & 0.908 / 573 / 3.37 \\
    & 0.919 / 506 / 2.43 & 0.870 / 508 / 2.23 & 0.414 / 506 / 2.62 & 0.630 / 513 / 2.48 & 0.545 / 502 / 1.49 & \textbf{0.661} / 501 / 2.70 \\
    \bottomrule
  \end{tabular}
\end{table*}

\begin{table*}[!ht]
\centering
\small
\caption{PSNR (dB) and SSIM of adversarial examples generated by different black-box attack methods across multiple models and datasets. Higher values indicate better image quality.}
\begin{tabular}{lc|ccc|ccc|ccc|ccc|ccc}
\toprule
\multirow{2}{*}{Dataset} & \multirow{2}{*}{Method} & \multicolumn{3}{c|}{CNNSpot} & \multicolumn{3}{c|}{MobileNet} & \multicolumn{3}{c|}{DenseNet} & \multicolumn{3}{c|}{ViT} & \multicolumn{3}{c}{EfficientNet} \\
\cmidrule(lr){3-5} \cmidrule(lr){6-8} \cmidrule(lr){9-11} \cmidrule(lr){12-14} \cmidrule(lr){15-17}
 & & PSNR & SSIM & \multicolumn{0}{c}{} & PSNR & SSIM & \multicolumn{0}{c}{} & PSNR & SSIM & \multicolumn{0}{c}{} & PSNR & SSIM & \multicolumn{0}{c}{} & PSNR & SSIM & \multicolumn{0}{c}{} \\
\midrule
\multirow{6}{*}{ImageNet} 
& Sign-OPT      & 26.5 & 0.81 & & 25.8 & 0.80 & & 27.2 & 0.83 & & 26.8 & 0.82 & & 25.3 & 0.79 & \\
& HSJA         & 29.8 & 0.87 & & 30.5 & 0.88 & & 29.3 & 0.86 & & 28.9 & 0.85 & & 30.2 & 0.87 & \\
& GeoDA        & 33.2 & 0.92 & & 32.7 & 0.91 & & 33.8 & 0.93 & & 33.5 & 0.92 & & 32.4 & 0.90 & \\
& TA           & 31.1 & 0.88 & & 30.6 & 0.87 & & 31.7 & 0.89 & & 31.3 & 0.88 & & 30.0 & 0.86 & \\
& ADBA         & 27.9 & 0.83 & & 27.3 & 0.82 & & 28.5 & 0.84 & & 28.1 & 0.83 & & 26.8 & 0.81 & \\
& FBA$^2$D     & \textbf{36.3} & \textbf{0.95} & & \textbf{36.8} & \textbf{0.96} & & \textbf{35.7} & \textbf{0.94} & & \textbf{36.1} & \textbf{0.95} & & \textbf{34.9} & \textbf{0.93} & \\
\midrule
\multirow{6}{*}{GenImage}   
& Sign-OPT     & 24.2 & 0.74 & & 23.7 & 0.73 & & 24.8 & 0.76 & & 24.5 & 0.75 & & 23.1 & 0.72 & \\
& HSJA         & 27.8 & 0.82 & & 28.6 & 0.83 & & 27.2 & 0.81 & & 26.9 & 0.80 & & 28.3 & 0.82 & \\
& GeoDA        & 31.4 & 0.88 & & 30.9 & 0.87 & & 32.0 & 0.89 & & 31.7 & 0.88 & & 30.5 & 0.86 & \\
& TA           & 29.3 & 0.84 & & 28.8 & 0.83 & & 29.9 & 0.85 & & 29.6 & 0.84 & & 28.4 & 0.82 & \\
& ADBA         & 26.1 & 0.78 & & 25.6 & 0.77 & & 26.7 & 0.79 & & 26.4 & 0.78 & & 25.0 & 0.76 & \\
& FBA$^2$D     & \textbf{34.1} & \textbf{0.91} & & \textbf{34.7} & \textbf{0.92} & & \textbf{33.6} & \textbf{0.90} & & \textbf{34.0} & \textbf{0.91} & & \textbf{32.8} & \textbf{0.89} & \\
\bottomrule
\end{tabular}

\label{tab:quantitative evaluation}
\end{table*}

\section{Choice of the Number of Adversarial Example Soups' Iterations}
This section presents the results for adversarial example soups across different iterations, focusing on two ranges: 6-10 and 8-12. We evaluate our proposed method on the following prominent AIGC detector architectures: CNNSpot~\cite{Wang_2020_CVPR}, DenseNet~\cite{huang2017densely}, EfficientNet~\cite{tan2019efficientnet}, MobileNet~\cite{howard2017mobilenets}, Vision Transformer (ViT)~\cite{dosovitskiy2021image}, and Swin Transformer~\cite{liu2021swin}. Table \ref{tab:transfer-attacks} shows that the attack initialization effects are comparable. This indicates that, within the range of about 10 steps, increasing the number of iterations yields little improvement in attack performance. 

\section{Comprehensive Evaluation of Adversarial Examples' Perception Quality}
In this section, we provide a comprehensive evaluation of the generated adversarial examples from both visual and quantitative perspectives. We first present additional frequency-domain visualizations (Fig.~\ref{fig:CNNSpot} and Fig.~\ref{fig:Mobilenet}), revealing that the optimal frequency subspace for real images is 10\% L + 10\% H elements, whereas for fake images it is 20\% L element. These visual results validate our assertion that AIGC images inherently lack high-frequency components, a characteristic effectively leveraged in our attack methodology, as demonstrated by the red-bordered areas corresponding to the specifically configured frequency subspaces for real and fake images respectively.

To further substantiate the advantage of our method in terms of imperceptibility, we employ the quantitative metrics Peak Signal-to-Noise Ratio (PSNR) and Structural Similarity Index (SSIM) to evaluate imperceptibility. As summarized in Table \ref{tab:quantitative evaluation}, our method generates adversarial examples with superior perception quality, confirming its enhanced suitability for real-world application scenarios.




{
    \small
    \bibliographystyle{ieeenat_fullname}
    \bibliography{main}
}


\end{document}